\documentclass[twocolumn]{aastex62}

\usepackage{graphicx}
\usepackage{dcolumn}
\usepackage{bm}
\usepackage{hyperref}
\usepackage{cancel}

\usepackage{blindtext}
\usepackage{braket}
\usepackage{booktabs}

\usepackage{wrapfig}
\usepackage{color}
\usepackage{amsmath}
\usepackage{amssymb}

\DeclareMathAlphabet\mathbfcal{OMS}{cmsy}{b}{n}
\DeclareMathAlphabet\mathcal{OMS}{cmsy}{n}{n}

\newcommand{\dd}{\mathrm{d}}

\def\aap{{Astronomy and Astrophys.}}

\def\apj{{\it Astrophys. J.\ }}
\def\apjs{{\apj\ \it Suppl.\ }}		
\def\apjl{{\apj\ \it Lett.\ }}

\submitjournal{ApJ}
\shorttitle{Steepening of Cosmic Ray Spectra}
\shortauthors{Hanusch et al.}
\begin{document}


\title{Steepening of Cosmic Ray Spectra in Shocks with Varying Magnetic
Field Direction}

\correspondingauthor{Adrian Hanusch}
\email{adrian.hanusch@uni-rostock.de}
\author{Adrian Hanusch}%
\affiliation{%
 Institut f\"ur Physik, Universit\"at Rostock, 18051 Rostock, Germany
}%
\author{Tatyana V. Liseykina}%
\affiliation{%
 Institut f\"ur Physik, Universit\"at Rostock, 18051 Rostock, Germany
}%
\affiliation{%
 Institute of Computational Mathematics and Mathematical Geophysics SB RAS, Lavrent'jev ave. 6, 630090 Novosibirsk, Russia}
\author{Mikhail Malkov}%
\affiliation{%
 CASS and Department of Physics, University of California, San Diego, La Jolla, California 92093, USA
}%
\author{Felix Aharonian}
\affiliation{Dublin Institute for Advanced Studies, 31 Fitzwilliam Place, Dublin
2, Ireland}

\affiliation{Max-Planck-Institut f\"ur Kernphysik, P.O. Box 103980, D 69029 Heidelberg,
Germany}

\date{\today}

\begin{abstract}
Cosmic ray (CR) spectra, both measured upon their arrival at the Earth's
atmosphere and inferred from the emission in supernova remnants (SNRs),
appear to be significantly steeper than the ``standard'' diffusive
shock acceleration (DSA) theory predicts. Although the reconstruction
of the primary spectra introduces an additional steepening due to
propagation effects, there is a growing consensus in the CR community
that these corrections fall short to explain the newest high-precision
data. Using 2D hybrid simulations, we investigate a new mechanism that may steepen the spectrum during the acceleration in SNR shocks.

Most of the DSA treatments are limited to  
homogeneous shock environments.
To investigate whether inhomogeneity effects can produce the necessary
extra steepening, we assume that the magnetic field changes its angle
along the shock front. The rationale behind this approach is the strong
dependence of the DSA efficiency upon the field angle, $\theta_\mathrm{Bn}$.
Our results show that the variation of shock obliquity along its face
results in a noticeable steepening of the DSA spectrum. Compared to
simulations of quasi-parallel shocks, we observe an increase of
the spectral index by $\Delta q=0.1-0.15$. Possible
extrapolation of the limited simulation results to more realistic
SNR conditions are briefly considered.
\end{abstract}


\section{Introduction}\label{sec:intro}
Cosmic ray (CR) spectra are now determined with  great accuracy revealing
noticeable disagreements with the theoretical predictions. Most of
these predictions rely on the diffusive shock acceleration theory
(DSA) -- a seemingly plausible long-standing solution for the production
of high-energy particles in a variety of shocks, primarily  
supernova remnant (SNR) shocks (see, \emph{e.g.}, \citealp{Schure2012} and references therein). 
The DSA theory, at least its basic version,
is remarkably insensitive to 
most shock parameters and predicts
a power-law spectrum in momentum, $f\propto p^{-q},$ for the
accelerated particles. Its index, $q=3r/\left(r-1\right)$, depends
only on the shock compression, $r$. Therefore, the DSA theory can be 
verified straightforwardly in strong
shocks with $r\approx4$. The agreement will then unlikely be coincidental,
but disagreement will raise doubts regarding the DSA as a viable explanation
for the observed spectrum. On the other hand, it would suggest to look
for missing elements in the ``standard'' DSA theory. These may still
reconcile the DSA even with high-precision observations.

Some difference between predicted and observed spectrum is expected
because of 
factors not determined by the experiments, primarily
propagation losses. Their scaling with energy is not known precisely,
so the tight DSA index can be increased (steepened) in the range $0.3-0.6$.
However, there are two types of observations where corrections for
the losses are not required to test the DSA predictions. The first
type is pertinent to the spectrum of the \emph{ratio} of two different
elements attributable to the same or similar accelerators.
At least in one such case, the theory has already faced a serious
challenge when a $\Delta q\approx0.1$ difference between the proton
and helium rigidity spectra was firmly established \citep{ATIC06,Adriani11,Aguilar15}.
Explanations have been given (see \citealp{Serpico2015,Malkov2018}
for a review), but only a few of them address the proton 
and helium spectra at the DSA level (see \citealp{Hanusch2019}
and the references therein). Even though they reproduce the data accurately, these 
explanations do not vindicate the
DSA completely because they obtain particle spectra  integrated   
over the entire history of an SNR evolution. Indeed,  the other type of observations 
where the disagreements are present
provide the momentary particle spectra. They
are measured \emph{in situ}, using secondary emission from SNRs, produced
by accelerated CR. 

At this point, it is worthwhile to separate the propagation effects
on the spectrum from the acceleration in the source. Only the latter
has a direct bearing on the DSA. At the same time, the CR propagation
models have long been used to offset for the DSA deviations from the
local spectra. A significant range in $\Delta q_{{\rm prop}}=0.3-0.6$,
depending on the particle scattering regime in the interstellar medium (ISM) (see, \emph{e.g.}, \citealp{Strong2007}), allowed one
to link the local spectrum measured on Earth, $E^{-2.7}-E^{-2.8}$ to that of the DSA-predicted, 
$E^{-2}-E^{-2.2}$. Being marginal in the first place,
this link became more questionable after tighter constraints have
been imposed on both the $\Delta q_{{\rm prop}}$ (by independent
measurements of the secondary to primary CR ratios, \citealp{Aguilar2016}) and the SNR source
spectra ($\gamma$-emission generated
by accelerated CRs through the $\pi^{0}$ channel, \emph{e.g.}, \citealp{Aharonian2019}).

Even more persuasive indication of a steep source spectrum is the recent high-precision  spectral index measured to be in the range $q=2.87\pm0.06 $ below 500 GeV, by the CALET team \citep{CALET_2019PhRvL.}, fully confirming the previously published AMS-02 proton spectrum \citep{Aguilar15}. These findings
have added to the CR community concern about the DSA capacity to account for the CR production in SNRs (e.g.,  \citealp{Gabici2019}). The old paradigm  explaining  steep spectra by imposing an exponential cut-off on the DSA-produced power-law does not comply with the new data, even if one assumes that the steep part of the spectrum originates from a nearby source with a cut-off in the sub-TeV range.     

To address the problem of spectrum steepening in the DSA
 several scenarios have been recently proposed. The authors of \citep{Bell2019} show that the steepened spectrum may be caused by the loss of CR energy due to turbulent magnetic field amplification during CR acceleration when the shock velocity is high.
Another particular scenario, 
\citep{Malkov_Aharonian2019}, 
utilizes a well-known property of collisionless shocks to suppress
the proton injection when the magnetic field ahead of the shock makes
a large angle to its normal, typically $\theta_\mathrm{Bn} \gtrsim \pi/4$.
Two specific steepening mechanisms can be associated with
the variable field inclination.

The first mechanism is best represented in bilateral SNRs, where two
regions of active particle acceleration ($\theta_\mathrm{Bn} \lesssim \pi/4$,
polar caps) are separated by a region of oblique shock geometry (equatorial
$\pi/4 \lesssim \theta_\mathrm{Bn} < \pi/2$) where no significant injection/acceleration
is observed. A clear-cut example here is SN 1006. The steepening mechanism
derives from the growth of the active acceleration zone in time, as
the SNR expands. Compared to the case of a fixed acceleration area,
freshly injected particles with lower energies are added to the growing
acceleration zone at an area-integrated rate that increases with time.
This enhanced production of low-energy particles naturally results
in a steeper overall spectrum \citep{Malkov_Aharonian2019}.

The second steepening mechanism 
is associated with particle diffusion from the active acceleration
zone into the neighboring oblique shock zone. As there is almost no
particle confining turbulence there, these particles have a good chance
to escape the accelerator. Since the diffusion coefficient typically
grows with energy, a steeper spectrum is expected. This mechanism
is much harder to quantify analytically since several competing factors
are at play. First, when a particle reaches the interface between
the active and inactive acceleration zones, the cross-field diffusion
decreases, whereas the parallel diffusion, on the contrary, increases.
This happens because the CR-driven turbulence level should decrease
at the interface, while the parallel and perpendicular particle diffusivities
are related as $\kappa_{\parallel}\kappa_{\perp}\simeq\kappa_{B}^{2}.$
Here $\kappa_{B}\propto r_{g}$ is the Bohm diffusion that linearly
growth with the particle gyro-radius $r_g.$  The turbulence decrease
will thus slow down the sideway losses but, at the same time, particles
reaching the interface are more likely to escape along the field.
If we add the wave self-generation by particles diffusing into the
``quiet'' zone of suppressed particle acceleration, the picture
becomes even more complicated for analytic treatment.

It may be noted from the preceding discussion that the second mechanism,
by contrast with the first one, does not require the growth of acceleration
area to produce steepening. Indeed, it results from particle losses
incurred at an energy-dependent rate. This suggests that the mechanism
can also work on a ``patchy'' shock where regions of active
and inactive acceleration are interspersed and coexist, not necessarily
growing or shrinking. Therefore, it should apply to planar shocks
and, to low-energy particles in spherical shocks where the shock curvature
is unimportant. This is the situation we study in this paper using
hybrid simulations with a magnetic field changing periodically along
the shock front.

It is important to note that the magnetic field ahead of the shock can vary
at sufficiently large scales due to the cyclotron instability of high
energy particles, escaping or diffusing further ahead of the shock.
In other words, the variable field component does not have to be preexisting
in the ISM. Under these circumstances, the shock geometry becomes
variable at scales significantly smaller than the shock radius but
still larger than the gyroradii of the low-energy accelerated particles.
The tilted magnetic field affects the injection of non-relativistic
particles. This makes hybrid simulations suitable for studying the
spectrum steepening mechanism outlined above, as no broad energy range
and, respectively, large simulation box is required.

In this paper, we study particle injection and acceleration in shocks
with magnetic field changing its direction along the shock face.
The simulation method is introduced in the next
section. Section~\ref{sec:Results} describes the principal results. We conclude with
a summary and brief discussion in Sec.~\ref{sec:Discussion}.

\section{Simulation setup}\label{sec:Set_up}
Hybrid simulations have been proven to be a valuable tool for investigating acceleration of ions at collisionless shocks in the solar system \citep{Lin2005,Burgess12,Giacalone2017} as well as CR injection and acceleration at SNR shocks \citep{Quest88, Caprioli2014}. In these simulations the electrons are considered as a massless, charge neutralizing fluid, while the ions are treated kinetically \citep[and references therein]{Lipatov2013}. In this paper we use a model, in which the electron pressure $p_{e}$ and the resistivity are both assumed to be isotropic, i.e. scalar quantities. 
The pressure $p_{e}$ is modeled using an adiabatic equation of state with 
the adiabatic index $\gamma_{e}=5/3.$ The fluid equations and the ion 
equations of motion are non-relativistic, as $|\vec{v}| \ll c$ holds during the injection phase.  More details on our hybrid code are given in \citep{Hanusch2019}.

In the simulations times are given in units of the inverse proton cyclotron frequency, $\omega_c = e B_0 / m_pc$, where $e, m_p$ are the protons charge and mass, 
respectively, $c$ denotes the velocity of light, and $B_0$ the amplitude of the background magnetic field. Distances are measured in terms of the proton skin depth 
$c/\omega_p$, with $\omega_p = \sqrt{4\pi \, n_0 \, e^2/m_p}$ being the proton plasma frequency,
 where $n_0$ is the far upstream density.
The velocity is given in terms of the Alfv\'en velocity 
$v_\mathrm{A} = B_0/\sqrt{4\pi \,n_0\,m_p}.$ 
Hence, the ratio of ion gyrofrequency to ion plasma frequency can be expressed as $\omega_c/\omega_p  =  v_\mathrm{A}/c$.
Spatially the motion of the ions is reduced to two dimensions, but all three components of the velocity and fields are included.

 Very recently the results of hydrodynamic modeling of the obliquity-dependent acceleration in a spherically expanding blast waves for different magnetic field 
morphologies have been reported \citep{Pais2018}. Earlier hybrid and fully kinetic particle-in-cell simulations of shocks with variable obliquity  
have been focused either on the interaction between the solar wind and a pla\-ne\-tary magnetic field \citep{Omidi2005,Blanco-Cano2009} or the simulations were initialized
with a radially expanding blast shell \citep{Dieckmann2018}. While the expanding blast shell setup resembles a spherically expanding SNR shock, it is computationally 
not feasible to follow its evolution over long enough timescales and observe the acceleration of ions to high energies.
In our simulation we mimic the variation of $\theta_\mathrm{Bn}$ over the shock surface in a 2D model setup. As the shock is propagating in the $x$-direction, 
we vary $\theta_\mathrm{Bn} = \theta_\mathrm{Bn}(y)$ as function of the transverse coordinate by properly choosing $B_x$ and $B_z$ as functions of $y$, ensuring $\nabla \vec B = 0$ and $|\vec B(y)| = B_0 = const$.

\begin{figure}
	\includegraphics[width=\linewidth]{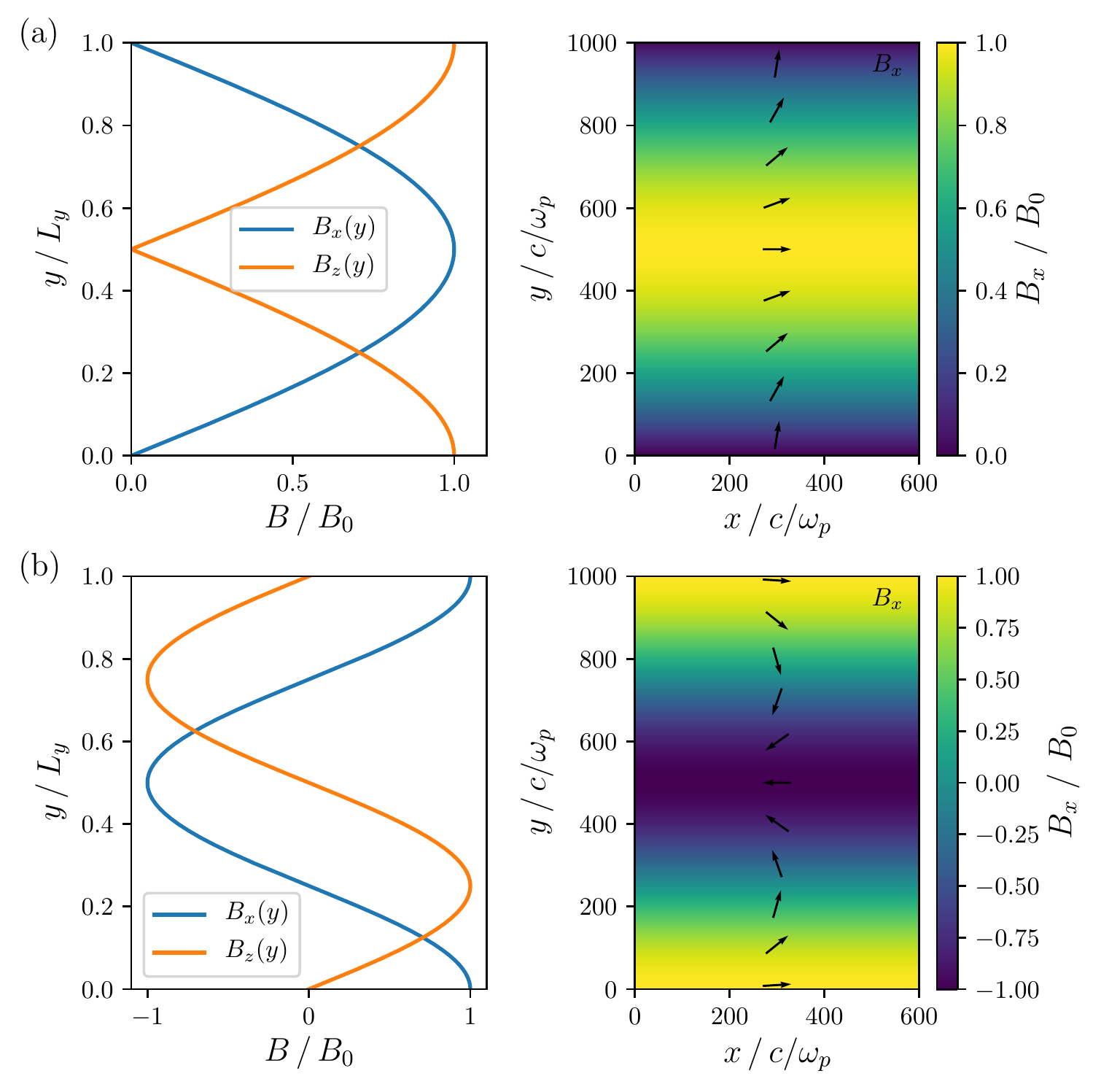}
	\caption{ Background magnetic field $\vec B_0(y)$ distribution at $t=0$ for two different
			  setups with varying shock obliquity.
			  The right panel shows initial dependence of the components $B_x$ and $B_z$ on the
			  transverse coordinate, while in the left panel the angle between the shock and the
			  background magnetic field is depicted by the black arrows in addition to 
			  $B_x(x,y)$.
			}
	\label{fig:setup}
\end{figure}

Two different setups of the initial magnetic field have been investigated. Both are depicted in Fig.~\ref{fig:setup}. While in the first setup the scale length of the magnetic field is twice the transverse size, in the second setup a full period is included in the simulation box. As both setups behave similarly we will focus on the first one when presenting the results.  
Note, that the magnetic field inhomogeneity upstream may be at different scales, starting from the shock radius of an SNR, for example, down to the resonant wave lengths of suprathermal (shock-reflected, injected, etc.)  particles. Our choice of the scale  is suitable to capture the physics of transition between the active and suppressed injection regimes. A smoother or spatially more complicated  magnetic field geometry would be interesting to explore, but we are constrained by the box size here.
The initial configuration of the electric field is calculated using the same algorithm as at later time steps. In particular, the electric field is obtained from the equation of motion of the massless electron fluid. Since the ions are initially distributed homogeneously across the grid the density is constant and the electron pressure term is zero. This setup warrants that the Lorentz force $\vec{F}_{\mathrm{L}}=q_i(\vec{E}+\vec{v}\times\vec{B})$ acting on the ions far upstream is zero. It has been tested for stability using a simulation box with periodic boundary conditions in $x$- and $y$-direction with a size of $L_x \times L_y = 1000 \times 1000 \;(c/\omega_p)^2$. The simulation showed no sign of instability on a time of $200 \, \omega_c^{-1}$. 

The simulation is initialized by sending a super-sonic and super-alfv\'enic hydrogen plasma  against a reflecting wall, placed  at $x=0.$ A shock forms upon the interaction of the 
counter-propagating plasma streams and propagates in positive $x$-direction. Thus, the simulation frame is the downstream rest frame. 

As the shock obliquity varies with the transverse coordinate, the transverse box size $L_y$ must be large enough, such that the results do not depend on $L_y.$ Our simulation box measures $L_x \times L_y = 8000 \times 1000 \; (c/\omega_p)^2,$ with the spatial resolution 
$\Delta_x=\Delta_y=0.5\;c/\omega_{p}.$ We use 16 particles per cell and a time step of $\Delta t = 0.001 \, \omega_c^{-1}$ for an initial upstream flow velocity of $v_0 = 10 \, v_\mathrm{A}$. Ions and electrons are assumed to be in thermal equilibrium far upstream with $\beta_e = \beta_p = 1.$  
The ion gyroradius $r_L = v_\perp/v_\mathrm{A} \;c/\omega_p$ ranges from 1 to 10 ion skin depths
depending on the magnetic field direction ($v_\perp = v_0$, where $\theta_\mathrm{Bn}=90^\circ$
and $v_\perp = v_\mathrm{th}$ for $\theta=0^\circ$.)
All numerical parameters have been checked for convergence. 

\section{Simulation results}\label{sec:Results}
\begin{figure}
	\includegraphics[width=1.02\linewidth]{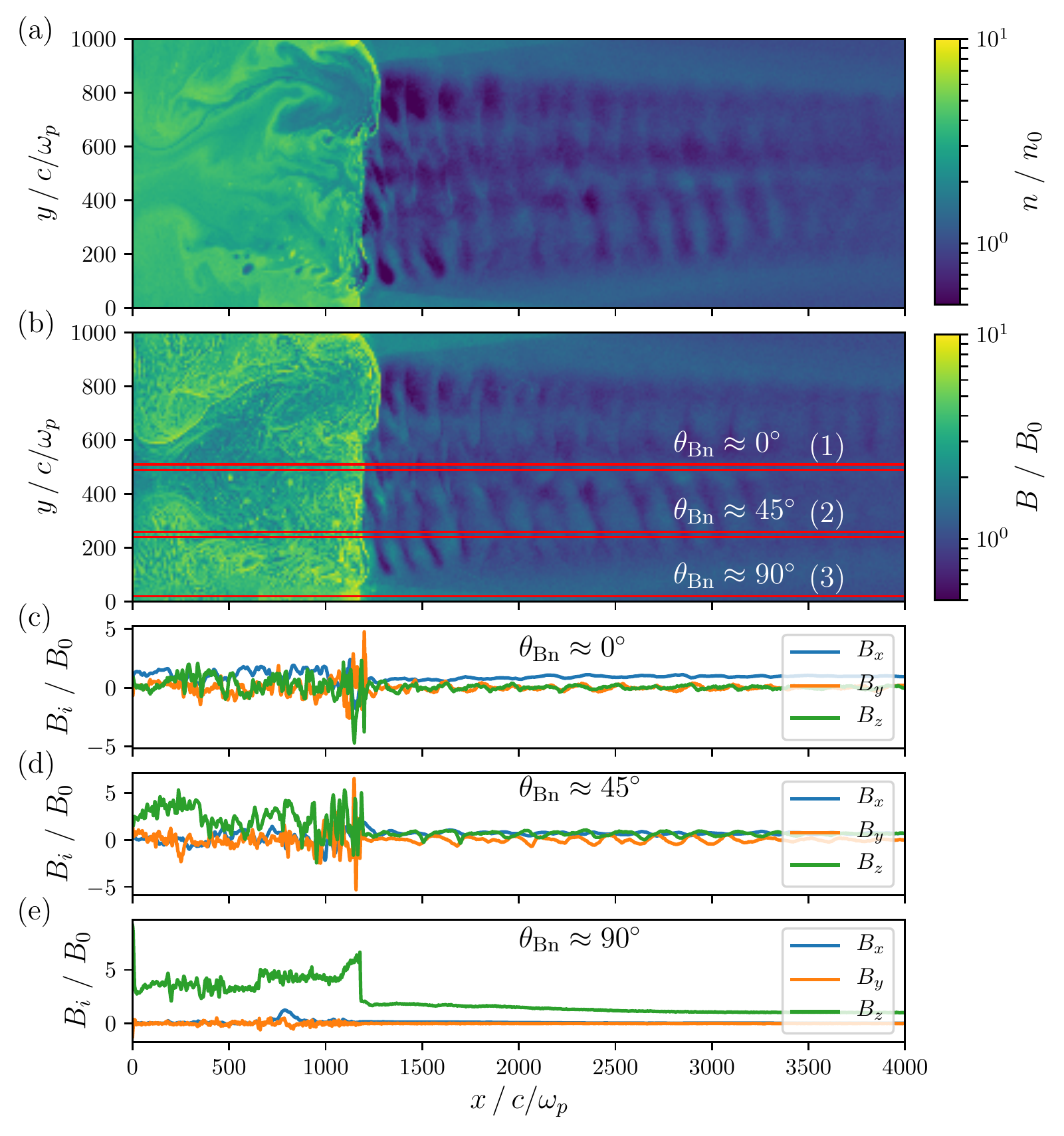}
	\caption{(a) -- Proton density $n(x,y)$  and (b) -- magnetic field amplitude $\vert \vec{B}\vert=
\sqrt{B_x^2+B_y^2+B_z^2}$ at $t=300\,\omega_c^{-1}$. The three panels (c)-(e) show the components of the magnetic field in regions of different shock obliquity, indicated by the red rectangles on frame (b).
			}
	\label{fig:fields}
\end{figure}

We start by presenting typical results from hybrid simulations of 
a collisionless shock propagating into an environment where the 
magnetic field inclination to the shock front varies. If not stated otherwise
the results correspond to an initial setup of the background magnetic field as 
depicted in Fig.~\ref{fig:setup}a).
Figure~\ref{fig:fields} shows a snapshot of the ion density $n(x,y)$ and the 
absolute value of the magnetic field $\vert \vec{B}(x,y)\vert$ in the simulation 
domain at $t=300\,\omega_c^{-1}.$ The shock transition at 
$x\sim 1200 \, c/\omega_p, $ the corresponding increase in density, and the 
compression of the magnetic field are clearly visible.
The three regions, indicated by the red rectangles on Fig.~\ref{fig:fields}b) 
correspond to the areas where the shock is parallel (1, $\theta_\mathrm{Bn} \approx 0^\circ$), quasi-parallel (2, $\theta_\mathrm{Bn} \approx 45^\circ$) or 
perpendicular (3, $\theta_\mathrm{Bn} \approx 90^\circ$). 
The components of the magnetic field in these regions 
(averaged over $y$) are shown on Fig.~\ref{fig:fields}c)-e).
The proton injection is only efficient at quasi-parallel shocks. This follows from simple kinematic considerations \citep{Malkov95} and is supported by Monte-Carlo \citep{Ellison1995} and hybrid simulations \citep{Caprioli2014}.
The angle beyond which the proton injection drops fast is close to $\theta_{\mathrm{cr}}=\pi/4.$ In fact, only in regions 
where the shock normal makes a reasonably sharp angle with the local magnetic field direction (quasi-parallel shock) particles can return upstream and 
excite waves there (frames (c) and (d)). In the quasi-perpendicular shock geometry region (Fig.~\ref{fig:fields}e) no waves are present in the upstream.

\begin{figure}
	\includegraphics[width=\linewidth]{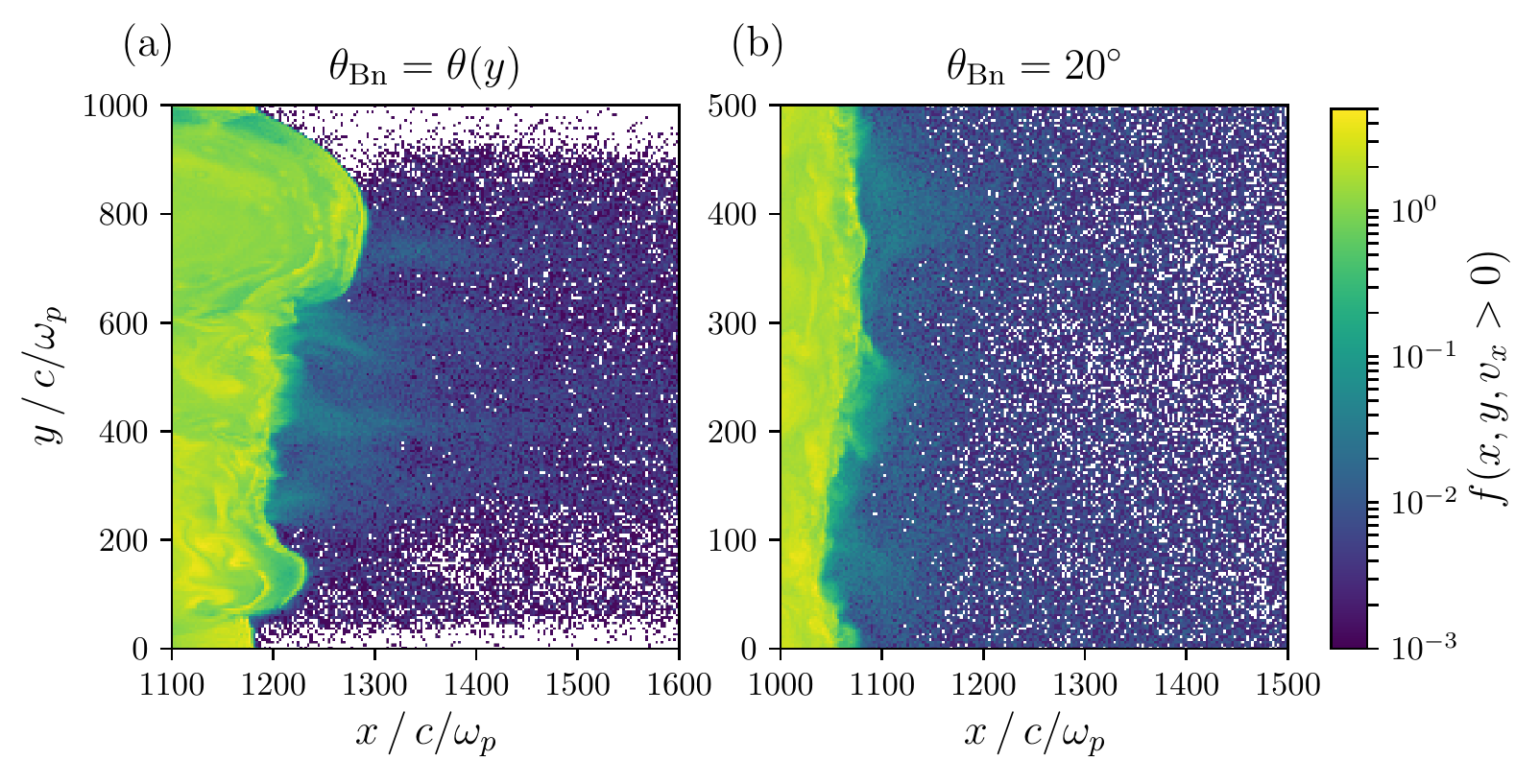}
	\caption{Density of protons with $v_x>0$ for (a) -- the model setup with variable obliquity
			 and (b) -- a simulation of a quasi-parallel shock, both at $t=300\,\omega_c^{-1}$. 
			 In the former case the presence of particles moving in $x$-direction in the upstream
			 is closely related to the condition for proton efficient injection 
			 $\theta_\mathrm{Bn} \le 45^\circ$.
			}
	\label{fig:positive_vx}
\end{figure}

\begin{figure}
	\includegraphics[width=\linewidth]{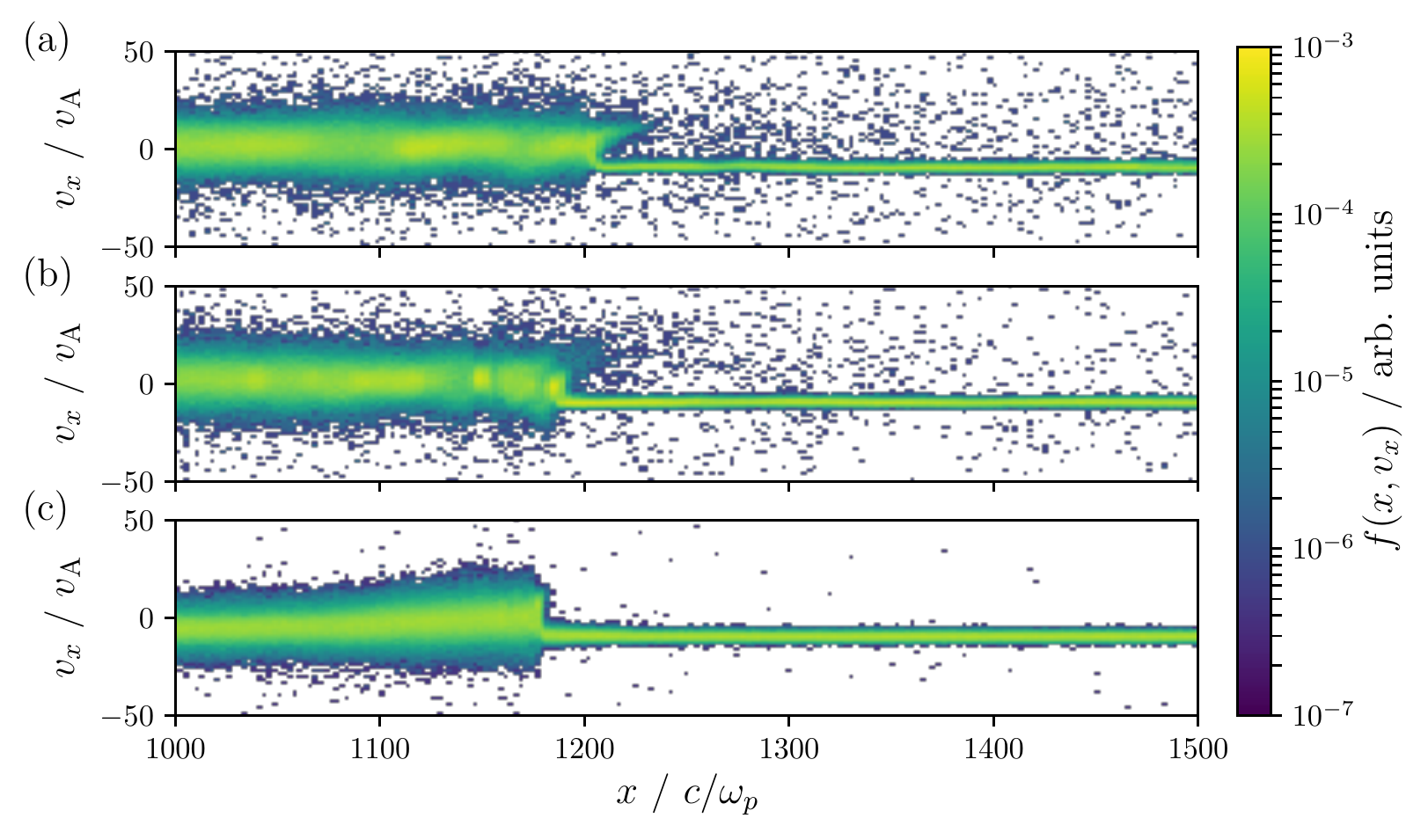}
	\caption{ Proton distribution on phase plane $f(x,v_x)$ at $t = 300 \, \omega_c^{-1}$ in the regions indicated in Fig.~\ref{fig:fields}b) by the red rectangles. For the particles in frame (a) the shock is parallel, for (b) -- quasi parallel and for (c) -- perpendicular.
			}
	\label{fig:phase_space}
\end{figure}

\begin{figure}
	\includegraphics[width=\linewidth]{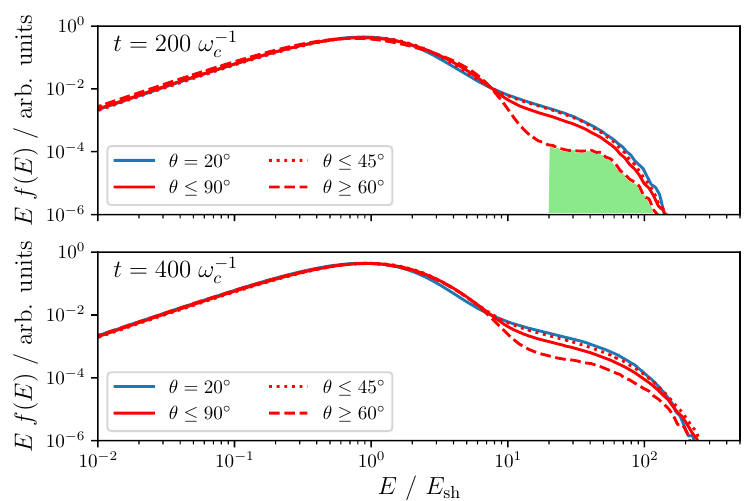}
	\caption{Downstream energy spectra: a simulation of a 
quasi-parallel shock (blue) and our model with variable shock obliquity (red) 
at two sequential time moments. Dotted (dashed) lines correspond to an 
integration over regions where the shock is quasi-parallel 
(quasi-perpendicular), (see Fig.~\ref{fig:setup}). The particles with $E > 20 E_\mathrm{sh}$ 
filling the green region at $t = 200\,\omega_c^{-1}$ are traced backwards in time in Fig.~\ref{fig:origin}.
			}
	\label{fig:spectra}
	\vspace{0.5cm}
\end{figure}

\begin{figure}
	\includegraphics[width=0.98\linewidth]{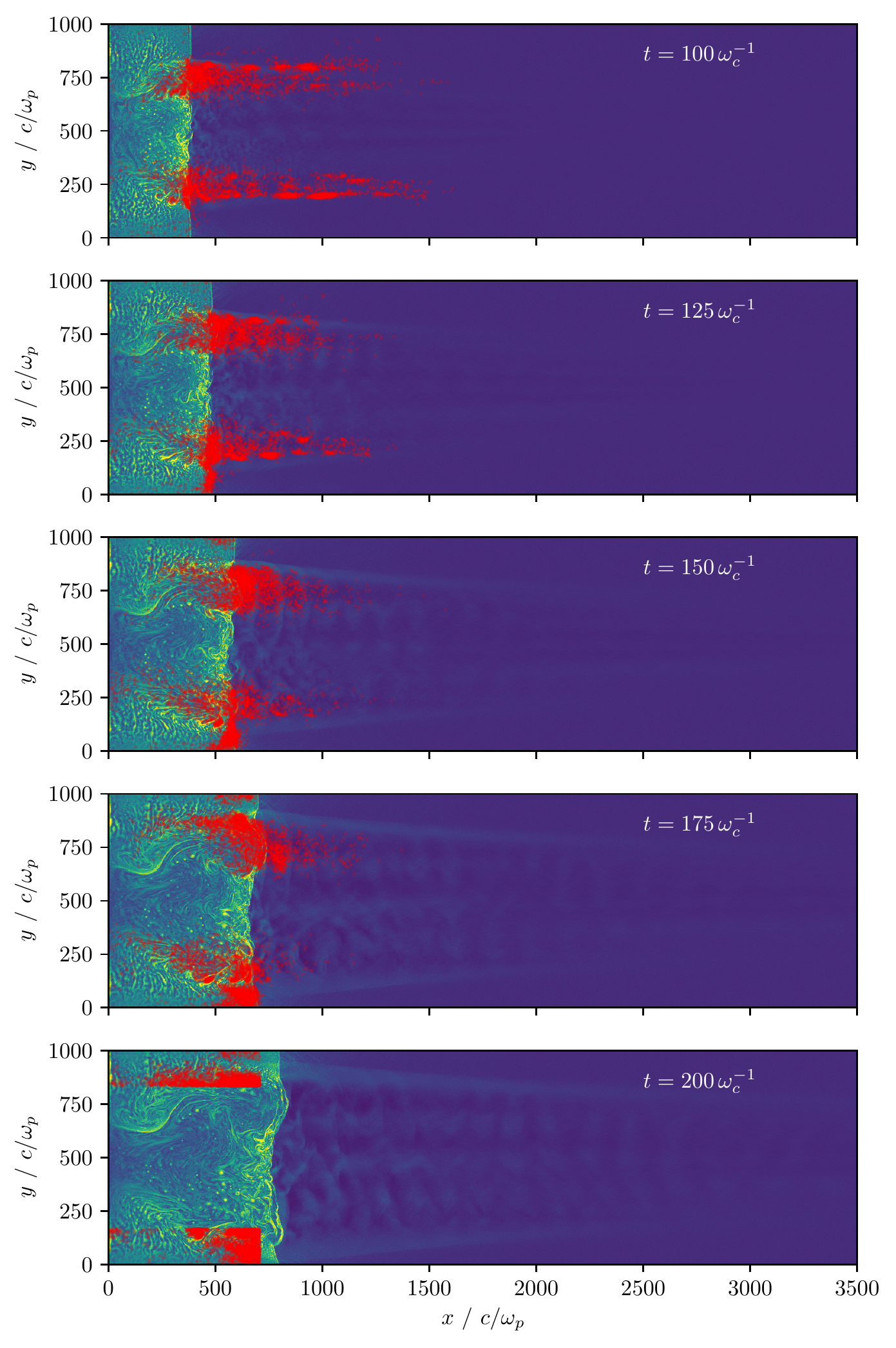}
	\caption{Distributions of accelerated particles (red dots) ending up at $t=200\,\omega_c^{-1}$  in the spectral tail $E > 20 E_\mathrm{sh}$ (marked green in Fig.~\ref{fig:spectra}) at earlier times. The background shows the distribution of the magnetic field amplitude at the corresponding times. 
			}
	\label{fig:origin}
\end{figure}

\begin{figure}[!h]
	\includegraphics[width=\linewidth]{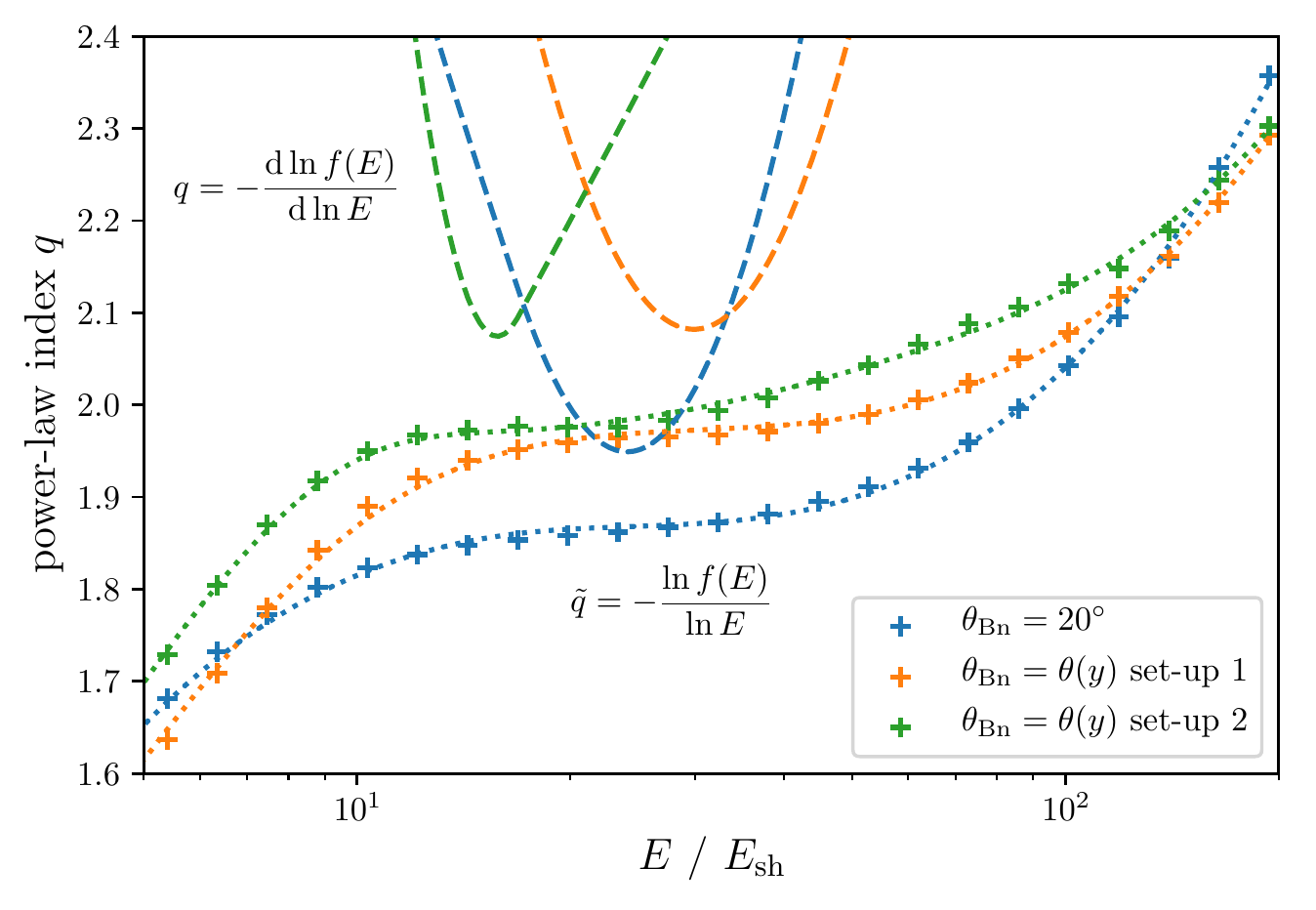}
	\caption{Power-law spectral index $q$
            calculated from the downstream energy spectra at $t = 400\,\omega_c^{-1}$: 
            blue line --  $\theta_\mathrm{Bn} = 20^\circ = const.,$ \
            orange line -- variable obliquity $ \vec{B}_0= B_0\left(\sin\frac{\pi y}{L_y}, 0, \left\vert\cos\frac{\pi y}{L_y}\right\vert\right), $ 
            green line -- variable obliquity $ \vec{B}_0= B_0\left(\cos\frac{2\pi y}{L_y}, 0,
\sin\frac{2\pi y}{L_y}\right).$ The discrepancy to the correct asymptotic index $q=1.5$ for $\theta_\mathrm{Bn} = 20^\circ$ is due to the limited run-time of the simulation and the limited box size. In the region closer to the shock particles are still accelerated to higher energies and a population of suprathermal particles exist in the transition region between the Maxwellian and the powerlaw. Our interest here is in the effect of shock obliquity which becomes evident already at $t = 400\,\omega_c^{-1}.$
	}
	\label{fig:exponent}
\end{figure}

The spatial distribution of protons with positive longitudinal 
velocity $v_x>0$ is shown in Fig.~\ref{fig:positive_vx}. Due to the 
turbulent fields in the downstream (note, that the simulation frame is 
the downstream rest frame), many particles have there a positive $v_x.$ On the contrary, only a relatively low amount of particles have positive $v_x$ in the upstream. These particles are either directly reflected from the shock or have returned from the downstream plasma and are mainly present in regions of $\theta_\mathrm{Bn} \le 45^\circ$ (Fig.~\ref{fig:positive_vx}a). For comparison the density of particles with positive $v_x$ is also shown for a shock with constant obli\-qui\-ty $\theta_\mathrm{Bn} = 20^\circ$ in Fig.~\ref{fig:positive_vx}b). In this case the number of particles with $v_x>0$ upstream of the shock is almost independent of the transverse coordinate and variations are only present close to the shock transition.\\
The proton distributions on the phase 
plane $f(x,v_x)$ in the regions of different shock 
obliquity (see Fig.~\ref{fig:fields}b) are shown in Fig.~\ref{fig:phase_space}.
They confirm the scenario above: shock 
reflected and accelerated particles originate from regions where the shock 
is quasi-parallel (Fig.~\ref{fig:phase_space}a,b). In the region where the 
shock is perpendicular (Fig.~\ref{fig:phase_space}c) the shock transition 
is very sharp and neither are reflected particles present in the upstream nor 
accelerated ions in the downstream. Furthermore, a slight difference of the 
shock positions for the parallel and perpendicular configuration becomes 
apparent. This can also be observed in the plot of the ion density 
(Fig.~\ref{fig:fields}a) in which the shape of the shock surface differs 
significantly from a planar one. 
However, we have not observed different shock velocities maintained over the whole time of the simulation for perpendicular and parallel regions.\\
In order to demonstrate the spectrum steepening effect, we have calculated proton energy 
spectra in the whole downstream, as well as in different obliquity regions (Fig.~\ref{fig:fields}b) for our model setup and compared them to the spectra obtained in  a simulation of a quasi-parallel shock with $\theta_\mathrm{Bn} = 20^\circ$. Figure~\ref{fig:spectra} shows the energy spectra at two different times, $t = 200\,\omega_c^{-1}$ and at $t=400\,\omega_c^{-1}$. The energy is given in terms of $\displaystyle E_\mathrm{sh} = m_p\, v_0^2/2$. It is to be expected that the energy spectra have a spatial dependence. \citet{Guo2010} have observed spatially dependent momentum distribution functions in simulations of collisionless shocks containing large-scale magnetic-field variations.
At early times the spectrum in the region of 
$\theta_\mathrm{Bn} \le 45^\circ$ of our model setup is similar to the spectrum, obtained in a simulation of a quasi-parallel shock. This is also not surprising  
since the injection efficiency does not depend largely on the angle between the shock normal and the background magnetic field, as long as the shock is quasi-parallel \citep{Caprioli2014}. By the end of the simulation, however, a deviation is visible as ions can move in the turbulent downstream plasma into regions, where the shock was initially quasi-perpendicular and vice versa. The energy spectrum in the quasi-perpendicular region exhibits a behavior which has been observed in simulations of perpendicular shocks: ions are mildly energized by shock drift acceleration \citep{Park2012} and the formation of a power-law tail is suppressed. But in contrast to fully (quasi-)perpendicular shocks the spectrum shows the presence of accelerated ions, which have diffused from the quasi-parallel region into the quasi-perpendicular one.
To confirm the origin of the accelerated particles we plot in Fig.~\ref{fig:origin} the positions of the particles ending up in the spectral tail $E > 20 E_\mathrm{sh}$ at $t = 200\,\omega_c^{-1}$ in the region $\theta_\mathrm{Bn} \ge 60^\circ$ (marked by the green region in the Fig.~\ref{fig:spectra})  at earlier times. The figure~\ref{fig:origin}  clearly shows, that the particles that contribute to the tail in the spectra in the quasi-perpendicular region ($\theta_\mathrm{Bn} > 60^\circ$) originate from the region where the shock normal is initially at an angle $\theta_\mathrm{Bn} < 60^\circ$ with respect to the background magnetic field (see panel $t = 100\,\omega_c^{-1}$ in Fig.~\ref{fig:origin}).\\
The energy spectra (Fig.~\ref{fig:spectra}) already indicate that the power-law exponent differs for the two simulation setups. To make it more evident, we have calculated the power-law index
\begin{equation}
	q = - \frac{\dd\ln\left( f(E) \right)}{\dd\ln E}
	\label{eq:power-law-index}
\end{equation}
from the computed spectra as function of energy and plot the result on 
Fig.~\ref{fig:exponent} (dashed lines). From the downstream energy spectra we have first 
calculated $\tilde q = \ln f(E) / \ln E$, indicated by the '+' markers and computed a spline (dotted line). This is then used to obtain $q$ according to Eq.~(\ref{eq:power-law-index}).
In the figure the blue lines corresponds to a setup with $\theta_\mathrm{Bn} = 20^\circ = const$, while the orange and green lines 
correspond to simulations with variable obliquity.
Common to all spectra is the beginning of the power-law around $E = 10\, E_\mathrm{sh}$ (see also Fig.~\ref{fig:spectra}) which has been reported as a result from earlier hybrid simulations. At high energies the spectra exhibit an exponential cut-off due to the limited simulation time and box size. 
In the region $10\, E_\mathrm{sh} < E < 50\, E_\mathrm{sh}$ $\ln f(E) / \ln E$ is almost constant and it is clearly visible, that the power-law exponent is larger for the simulation setups with variable shock obliquity. The difference of the spectral indices obtained from the simulations is $\Delta q = q_\mathrm{var} - q_\mathrm{const} \approx 0.1-0.15$.\\

\section{Summary}\label{sec:Discussion}
We have investigated the influence of a variable orientation
of the upstream magnetic field on ion acceleration at collisionless shocks
using hybrid simulations.
However limited in resources, our simulations have captured a new
physical phenomenon in the DSA -- the spectrum steepening associated
with the variation of shock obliquity along its face. The principal
results of this paper are as follows.\\
(i) For regions of parallel and perpendicular shock propagation coexisting
in one simulation box the DSA process, as expected, starts and proceeds
at high efficiency in the quasi-parallel region.\\[1mm]
(ii) 
However, while the particle energy increases, some of them diffuse
to the quasi-perpendicular shock domain.\\[1mm]
(iii) The resulting shock-generated downstream spectrum is noticeably steeper
than canonical DSA result obtained for a quasi-parallel shock.\\[1mm]
(iv) No strong shock deceleration in the quasi-perpendicular field direction
compared to the quasi-parallel one is observed.

While the spectrum steepening observed in our simulations may be regarded
as relatively small, $\Delta q=0.1-0.15$,  it is significant for a
number of reasons. First, it is likely to be scalable to larger boxes
and longer simulation time, so that somewhat larger $\Delta q$ values
can be expected for realistic SNR conditions. Second, the mechanism
operates at relatively low energies while at higher energies a different
but closely related mechanism, described in the Introduction, should
take over. Again, a steeper overall spectrum in the SNRs will then
result, as suggested by observations. Finally, the highly determined
DSA predictions combined with extremely accurate CR data have proved
even small variations of the spectral index to be meaningful and even
testable. The $\Delta q\approx0.1$ difference between the helium
and proton local spectra is a good example.

\begin{acknowledgments}
The research was supported by the DFG within the Research grant 278305671 and in part by 
NASA ATP-program within grant 80NSSC17K0255 and by the National Science Foundation under Grant No. NSF PHY-1748958. TVL acknowledges support of the Russian Science Foundation through the grant 19-71-20026 in the part related to the development of the analytic model and analysis of numerical results. Simulations were performed using the computing resources granted by the North-German Supercomputing Alliance (HLRN) under the project mvp00015. 
\end{acknowledgments}

\bibliography{literature}

\end{document}